\documentclass[twoside,a4paper,11pt]{sea10}
\usepackage{graphicx}
\usepackage{hyperref}
\usepackage{movie15}
\begin{document}
\pagenumbering{arabic}
\pagestyle{myheadings}
\thispagestyle{empty}
{\flushleft\includegraphics[width=\textwidth,bb=58 650 590 680]{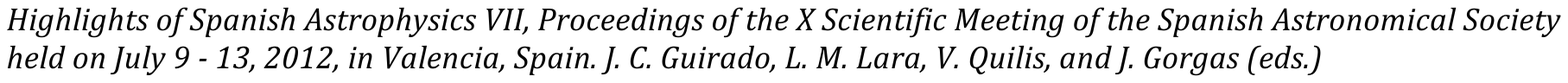}}
\vspace*{0.2cm}
\begin{flushleft}
{\bf {\LARGE
%
Discovery and spectroscopic study of the massive Galactic cluster Mercer 81.
%
}\\
\vspace*{1cm}
%
Diego de la Fuente$^{1}$,
Francisco Najarro$^{1}$, 
Ben Davies$^{2}$
and
Donald F. Figer$^{3}$
%
}\\
\vspace*{0.5cm}
%
$^{1}$
Centro de Astrobiolog\'ia (CSIC/INTA), ctra. de Ajalvir km. 4, 28850 Torrej\'on de Ardoz, Madrid, Spain\\
$^{2}$
Institute of Astronomy, University of Cambridge, Madingley Road, Cambridge CB3 0HA, UK\\
$^{3}$
Center for Detectors, Rochester Institute of Technology, 54 Lomb Memorial Drive, Rochester, NY 14623, USA
%
\end{flushleft}
%
\markboth{
The massive Galactic cluster Mercer 81
}{ 
%
D. de la Fuente et al.
%
}
\thispagestyle{empty}
\vspace*{0.4cm}
\begin{minipage}[l]{0.09\textwidth}
\ 
\end{minipage}
\begin{minipage}[r]{0.9\textwidth}
\vspace{1cm}
\section*{Abstract}{\small
%

During the last decade, hundreds of young massive cluster candidates have been detected in the disk of the Milky Way. We investigate one of these candidates, Mercer 81, which was discovered through a systematic search for stellar overdensities, with follow-up NICMOS/HST infrared narrow-band photometry to find emission-line stars and confirm it as a massive cluster. Surprisingly, the brightest stars turned out to be a chance alignment of foreground stars, while a real massive cluster was found among some fainter stars in the field. From a first spectroscopic study of four emission-line stars (ISAAC/VLT), it follows that Mercer 81 is a very massive young cluster, placed at the far end of the Galactic bar. Additionally, in this work we present some unpublished spectra from a follow-up observation program which confirm that the cluster hosts several Nitrogen-rich Wolf-Rayet stars (WN) and blue supergiants.

%
\normalsize}
\end{minipage}
%
%
%
\section{Introduction}

Since the 1990s, infrared astronomy has experienced a considerable development, allowing to observe in detail the most extincted regions of the Milky Way. As a consequence, many obscured young massive clusters (YMCs) have been found in the Galactic disk, revealing a great amount of previously unknown star formation. YMCs are notable for hosting a population of massive stars, which typically consists of OB, Wolf-Rayet and hypergiant stars. YMCs are also ideal test beds to measure the high-mass region of the IMF (\cite{imf}) and the evolution of massive stars (\cite{evolmassive}), as well as useful tools to map the Galactic metallicity (\cite{diskmetals}).

Only a few Galactic YMCs have been investigated extensively, e.g. Westerlund 1 (\cite{wd1}), the Arches Cluster (\cite{arches}) or the Quintuplet Cluster (\cite{quintuplet}). However, several systematic searches for IR clusters have been carried out (\cite{survey1}, \cite{survey3}, \cite{survey4}), yielding hundreds of new cluster candidates that mostly remain unstudied. The usual way to confirm such candidates as real YMCs consists of finding their young massive stars, whose high mass-loss rates allow to track them as highly reddened emission-line stars, especially by means of their Paschen-$\alpha$ emission excess (\cite{paschena}). A subsequent spectroscopic analysis of these stars eventually lead to a complete characterization of each cluster.

\section{Discovery and rediscovery \label{discovery}}

\begin{figure}
\center
\includegraphics[width=13.5cm]{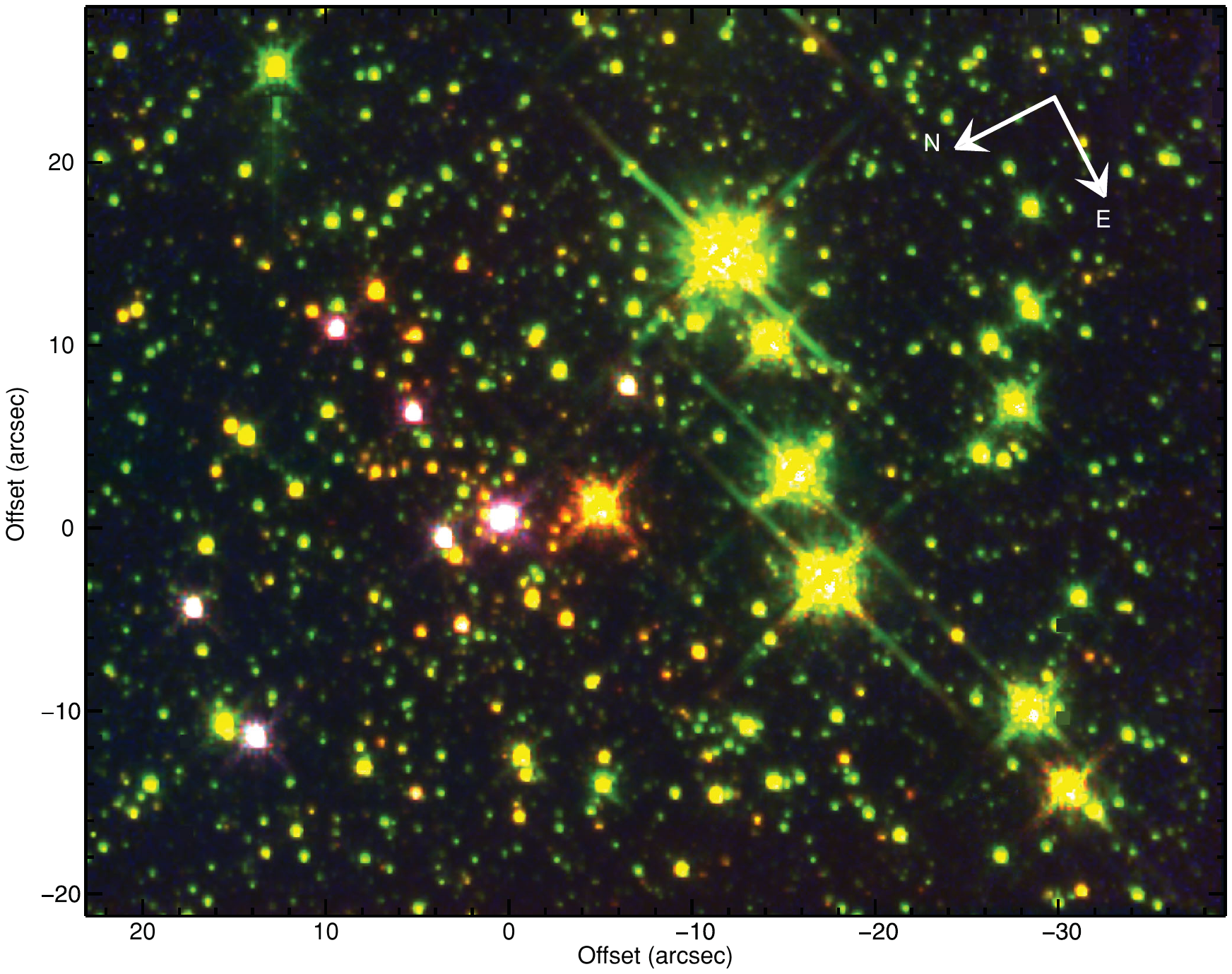}
\caption{\label{fig1} RGB image of Mercer 81, composed of NICMOS/HST photometric data as follows: F222M (red), F160W (green) and [F187N$-$F190N] (blue). From \cite{mc81}.
}
\end{figure}

The cluster candidate Mercer 81 was found in 2005 by \cite{survey2} in an algorithmic search for stellar overdensities in the GLIMPSE point-source catalog. In 2008, this candidate was observed with the instrument NICMOS onboard the Hubble Space Telescope as part of the observing program \#11545, whose main goal was to find highly reddened emission-line stars in candidate clusters. The strategy consisted of obtaining images through the filters F160W and F222M in order to measure the reddening; as well as narrow-band images at the wavelengths of Paschen-$\alpha$ (F187N) and the continuum region near $P_\alpha$ (F190N), in such a way that the subtraction image F187N$-$F190N would pinpoint the massive stars.

\begin{figure}
\center
\includegraphics[width=15.5cm]{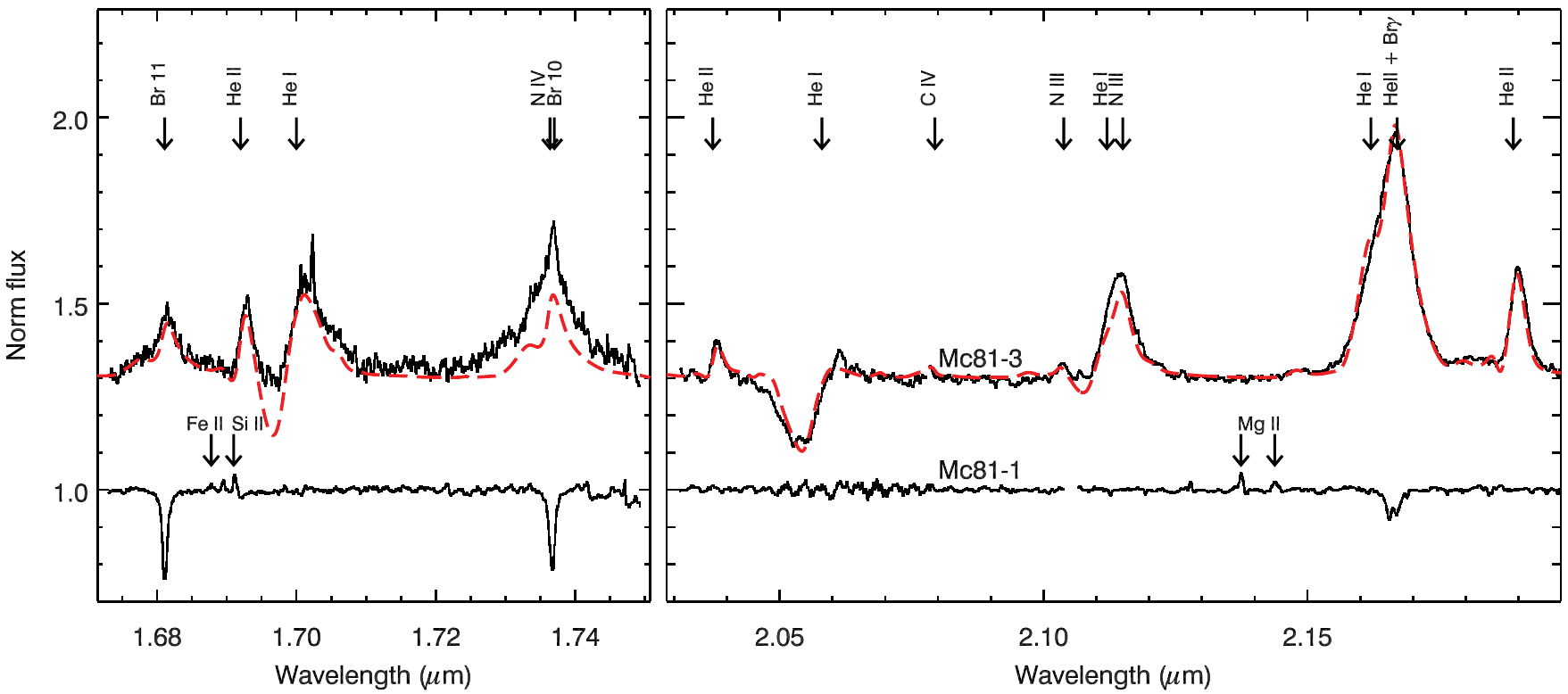}
\caption{\label{fig2} The two only spectra that could be completed in the first observation program at ISAAC/VLT. From \cite{mc81}.
}
\end{figure}

Results of the aforementioned photometry were published by \cite{mc81}, who presented a composite false-color image of Mercer 81 (Fig.~\ref{fig1}; see caption for the RGB-color explanation) where stars can be easily identified by means of their color. The most reddened stars must present red/orange colors, while the $P_\alpha$ excess corresponds to a blue color; therefore, the massive members of the cluster are expected to have a combination of these colors, appearing pink/magenta sources. On the other hand, the unreddened foreground stars look yellow/green. Fig.~\ref{fig1} shows that the brightest stars in the field constitute a chance alignment of foreground stars, which probably were crucial for the candidate detection. However, a fainter but numerous group of highly reddened stars, including nine emission-line stars, can be seen in the north half of the field

\section{Spectroscopy}

In 2009, H- and K-band spectroscopy of the emission-line stars at ISAAC/VLT was proposed (program ID: 083.D-0765); unfortunately, due to a temporary failure of the instrument, only the spectra of two cluster members could be completed. These spectra (Fig.~\ref{fig2}), which consisted of a late-B/early-A supergiant and a Nitrogen-rich Wolf-Rayet star (spectral type: WN7-8), have been published by \cite{mc81}, where a model for the WN achieved with the CMFGEN code (\cite{cmfgen1}, \cite{cmfgen2}) was included.

\begin{figure}[t!]
\center
\includegraphics[width=15.5cm]{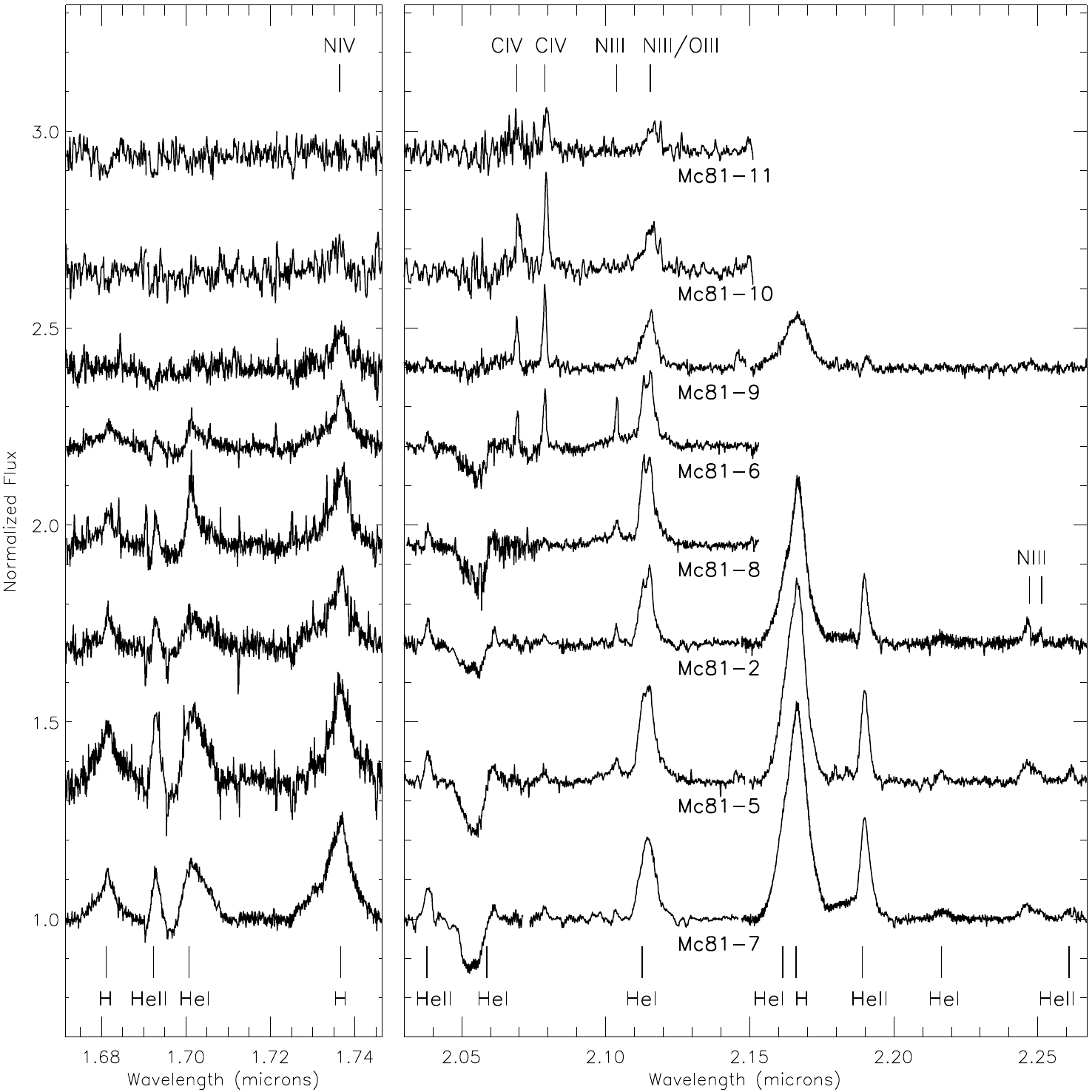}
\caption{\label{fig3} New H- and K-band spectra of Mercer 81 cluster members. Objects follow the naming scheme of \cite{mc81}, with the addition of new cluster member Mc81-11. 
}
\end{figure}

In 2011, we could complete the observations with additional spectra of 8 emission-line stars at ISAAC/VLT (program ID:087.D-0957), which are presented in Fig.~\ref{fig3}. All these objects show H and He broad emission lines, indicating extended winds and high mass-loss rates.

Three of the new spectra (Mc81-2, Mc81-5 and Mc81-7) are almost identical to Mc81-3, therefore having the same spectral type (WN7-8). Three others (Mc81-6, Mc81-9 and Mc81-10) are intermediate-O supergiants, as clearly showed by He\textsc{ii} features along with narrow C\textsc{iv} emission lines. Spectrum Mc81-11 might be as well classified as a O supergiant, although the low signal-to-noise ratio does not allow to confirm it. Finally, spectrum Mc81-8 shows intermediate features between a O supergiant and WN, perhaps entailing this object is in transition between these evolutionary stages (\cite{transition}).

\section{Discussion and future work}

Based on photometry (section~\ref{discovery}) and spectroscopic analysis of the 2 firstly observed spectra, \cite{mc81} presented a first characterization of Mercer 81 that now can be improved by means of the more complete spectroscopic data presented here. Particularly, the total mass of Mercer 81 (a few $\times 10^4$) given by \cite{mc81} was estimated assuming that all the unobserved emission-line stars were WN and that the cluster has the same evolutionary stage than Westerlund 1 (\cite{wd1evol}). However, our new data clearly suggest that Mercer 81 is younger, as we have detected earlier spectral types (O5-6 I) with respect to Westerlund 1 (O9 I, maximum), as well as the presence of emission-line stars other than WNs.

Since we are currently in the process of modeling the observed spectra, we cannot present here definitive results yet. Ongoing NLTE spherical atmosphere models will yield accurate spectral types and measure stellar and wind properties, including chemical abundances. This will result in a complete characterization of Mercer 81 which could be crucial to understand the chemodynamics of the inner disk due to its privileged location, at the far end of the Galactic Bar (\cite{mc81}). On the other hand, the extreme and uncommon objects belonging to this cluster apparently form an evolutionary sequence that may turn Mercer 81 into an ideal laboratory to study the final stages of massive stars.

%
%
\small  
%
\section*{Acknowledgments}   
%
This research was partially supported by the Ministerio de Ciencia e Innovaci\'on through grants AYA2008-06166-C03-02 and AYA2010-21697-C05-01. D.F. also acknowledges financial support from the FPI-MICINN predoctoral fellowship BES-2009-027786.

%

%

\begin{thebibliography}{}
\small
%
\bibitem{survey4}{Borissova, J., Bonatto, C., Kurtev, et al. 2011, A\&A, 532, 131}
\bibitem{wd1evol}{Brandner W.  Clark, J.S., Stolte, A., et al. 2008, A\&A, 478, 137}
\bibitem{wd1}{Clark, J. S.; Negueruela, I., Crowther, P. A., 2005, A\&A, 434,949}
\bibitem{mc81}{Davies, B., de la Fuente, D., Najarro, F., et al. 2012, MNRAS, 419, 1860}
\bibitem{survey1}{Dutra, C.M., \& Bica, E., 2001, A\&A, 676, 434}
\bibitem{imf}{Espinoza, P., Selman, F.J., \& Melnick, J., 2009, A\&A, 501, 563}
\bibitem{quintuplet}{Figer, D.F., McLean, I.S., \& Morris, M., 1999, ApJ, 514, 202}
\bibitem{arches}{Figer, D.F., Najarro, F., Gilmore, D., et al. 2002, ApJ, 581, 258}
\bibitem{survey3}{Froebich, D., Scholz, A., \& Raftery, C.L., et al. 2007, MNRAS, 374, 399}
\bibitem{cmfgen1}{Hillier, D.J., 1989, A\&A, 231, 116}
\bibitem{cmfgen2}{Hillier, D.J., \& Miller, D.L., 1998, ApJ, 496, 407}
\bibitem{diskmetals}{Najarro, F., 2008, IAUS, 250, 265}
\bibitem{transition}{Morris, P.W., Eenens, P.R.J., Hanson, M.M., Conti, P.S. \& Blum, R.D., 1996, ApJ, 470, 597}
\bibitem{evolmassive}{Martins, F., Genzel, R., Hillier, D.J., et al. 2007, A\&A, 468, 233}
\bibitem{survey2}{Mercer, E.P., Clemens, D.P., Meade, M.R., et al. 2005, ApJ, 635, 560}
\bibitem{paschena}{Wang, Q.D., Dong, H., Cotera, A., et al. 2010, MNRAS, 895, 902}

%
%
\end{thebibliography}
\end{document}